\documentclass[10 pt, conference]{IEEEtran}  

\usepackage{amsmath} 
\usepackage{bm}
\usepackage{amssymb}  
\usepackage{xcolor}
\usepackage{graphicx}
\usepackage{subcaption}
\usepackage{cite}
\usepackage[english]{babel}
\usepackage{blindtext}
\usepackage{lipsum}
\usepackage{amsthm}
\usepackage{dsfont}
\usepackage[linesnumbered, ruled]{algorithm2e}
\SetKwRepeat{Do}{do}{while}%

\usepackage{pgfplots}
\pgfplotsset{compat=newest}
\usepgfplotslibrary{groupplots}
\usepgfplotslibrary{dateplot}
\usetikzlibrary{calc,intersections,through,backgrounds,shapes,decorations.pathmorphing,arrows,plotmarks,external,patterns}
\tikzsetexternalprefix{./tikz/externaltikz/}

\usepackage[absolute,showboxes]{textpos}

\TPMargin{5pt}

\newcommand{\copyrightstatement}{
	\begin{textblock}{0.84}(0.08,0.01)    
		\noindent
		\footnotesize
		\copyright 2019 IEEE. Personal use of this material is permitted. Permission from IEEE must be obtained for all other uses, in any current or future media, including reprinting/republishing this material for advertising or promotional purposes, creating new collective works, for resale or redistribution to servers or lists, or reuse of any copyrighted component of this work in other works.
	\end{textblock}
}

\title{Optimal Scheduling for Discounted Age Penalty Minimization in Multi-Loop Networked Control}

\author{
	\IEEEauthorblockN{Onur Ayan, Mikhail Vilgelm, Wolfgang Kellerer}
	\IEEEauthorblockA{Chair of Communication Networks}
	\IEEEauthorblockA{Technical University of Munich, Germany}
	\IEEEauthorblockA{\{onur.ayan, mikhail.vilgelm, wolfgang.kellerer\}@tum.de}
}

\begin{document}
\copyrightstatement
\thispagestyle{empty}
\pagestyle{empty}

\maketitle
\definecolor{myred}{RGB}{220,43,25}
\definecolor{mygreen}{RGB}{0,146,64}
\definecolor{myblue}{RGB}{0,143,224}
\definecolor{mygray}{gray}{0.90}

\newif\ifcomments

\commentstrue
\newcommand{\commentBy}[3]{\textcolor{#1}{\{#2: #3\}}}
\newcommand{\mv}[1]{\ifcomments\commentBy{myblue}{MV}{#1}\fi}
\newcommand{\oa}[1]{\ifcomments\commentBy{myred}{OA}{#1}\fi}

\newtheorem{mydef}{Definition}
\newcommand{\setOfULResources}{\mathcal{R}^{\text{UL}}}
\newcommand{\setOfDLResources}{\mathcal{R}^{\text{DL}}}
\newcommand{\numOfULResources}{\mathsf{R^{\text{UL}}}}
\newcommand{\numOfDLResources}{\mathsf{R^{\text{DL}}}}
\newcommand{\ulResourceConsumption}{r_i^{\text{UL}}}
\newcommand{\dlResourceConsumption}{r_i^{\text{DL}}}
\newcommand{\setOfSamples}{\Xi_i}
\newcommand{\txDelay}{d}
\newcommand{\numLoops}{N}
\newcommand{\bs}{\text{BS}}
\newcommand{\avgAge}{\overline{\Delta}}
\newcommand{\iae}{\Sigma_{e}}
\newcommand{\samplingPeriod}{T^s_i}
\newcommand{\E}{\mathbb{E}}
\newcommand{\tr}{\mathsf{tr}}
\newcommand{\ki}{k_i}
\newcommand{\plant}{\mathcal{P}_i}
\newcommand{\sensor}{\mathcal{S}_i}
\newcommand{\controller}{\mathcal{C}_i}
\newcommand{\estimator}{\mathcal{E}_i}
\newcommand{\errornorm}{\E\left[\left\Vert e_i[k] \right\Vert^2 \right]}
\newcommand{\statespace}{\bm{\mathcal{\bm{S}}}}
\newcommand{\approxstatespace}{\bm{\mathcal{S}}_M}

\begin{abstract}	
Age-of-information (AoI) is a metric quantifying information freshness at the receiver. Since AoI combines packet generation frequency, packet loss, and delay into a single metric, it has received a lot of research attention as an interface between communication network and application.
In this work, we apply AoI to the problem of wireless scheduling for multi-loop networked control systems (NCS), i.e., feedback control loops closed over a shared wireless network. We model the scheduling problem as a Markov decision process (MDP) with AoI as its observable states and derive a relation of control system error and AoI. We further derive a stationary scheduling policy to minimize control error over an infinite horizon. We show that our scheduler outperforms the state-of-the-art scheduling policies for NCS. To the best of our knowledge, this is the first work proposing an AoI-based wireless scheduling policy that minimizes the control error over an infinite horizon for multi-loop NCS.
\end{abstract}
\section{INTRODUCTION}
Cyber-Physical Systems (CPS) are engineered systems that integrate computation and control with physical processes. Power grids, smart transportation, and industrial robotics are some prominent applications of CPS~\cite{lee2008cyber}. From a system theoretical point of view, many CPS fall under the category of Networked Control Systems (NCS), i.e., feedback control loops closed over a communication network. Each loop consists of a plant, a sensor, and a controller with state estimator. In a typical NCS scenario, multiple heterogeneous NCS share a wireless communication medium and compete for network resources to transmit their latest sensor measurements to the controllers. The received measurement is then used to estimate the plant state remotely and determine the best control input for the plant. In such a setting, imperfections of the wireless channel or time-critical requirements of the underlying control loops impose challenges for both communication and control. In particular, the end-to-end delay or information loss induced by the network reduces the precision of control actions, hence degrading the control quality. 

Age-of-information (AoI) is a metric that measures the information freshness about a remote process at the receiver side. It is defined as the time elapsed since the generation of the latest received information \cite{kaul2012real, kaul2012status}. As the name suggests, AoI increases linearly in time for all types of applications until a new update is received. AoI combines packet generation frequency, end-to-end delay, and packet loss in a single metric. For example, the absence of information increases AoI on the controller side irrespective of the cause: high delay, packet loss, or a low information update frequency. Due to its ability to connect control and communication layers, AoI has been widely adopted as a cross-layer metric by the researchers from both communities. 

\subsection{Related Work}
The initial research interest on AoI has concentrated solely on the communication side. The effect of various queuing and packet management techniques has been studied for single-hop  \cite{kaul2012real, kaul2012status, costa2014age, huang2015optimizing, sun2017update} and multi-hop \cite{bedewy2017age, talak2017minimizing} transmissions. \cite{kadota2016minimizing, talak2017minimizing, hsu2017age} study the AoI minimization as a resource scheduling problem and provide average optimal policies, i.e., policies that are optimal to achieve minimum average AoI in the network. The notion of age penalty functions, that is, staircase or non-linear evolution of AoI over time, has been proposed to reflect the heterogeneous demands of different applications sharing a communication network as in typical industrial internet of things or CPS scenarios. \cite{sun2017update} conducts a study on update policies for age penalty minimization in the presence of random transmission delays. They consider a single-user link and show that generating a new information right after the delivery of the previous packet is not optimal under certain conditions. In \cite{kosta2017age}, authors expand the concept of AoI by introducing Cost of Update Delay and Value of Information of Update Delay metrics as a function of information staleness. They consider single-user link with first-come-first-served (FCFS) queue discipline for a M/M/1 model.

With the introduction of application-dependent age penalty functions, the AoI metric has been adopted in NCS research as well. In \cite{ayan2019age} authors consider multiple heterogeneous feedback control loops communicating over a base station where the network resources are limited. They define estimation error as a function of AoI and compare age- and value of information (VoI) scheduling of wireless resources. In the context of NCS, they show that optimal AoI scheduling leads to worse estimation accuracy than greedy VoI scheduling. Similarly, in \cite{klugel2019aoi} authors study the estimation problem for stochastic processes in a single-loop and single-hop scenario. They propose average optimal sampling strategies for age penalty minimization for selected age and transmission cost functions. \cite{vilgelm2017control} proposes a heuristic scheduling policy for wireless NCS where network resources are allocated greedily to the sub-systems with the highest expected error.

\subsection{Contributions}
The key contribution of this work is the derivation of an approximately optimal scheduling strategy for a \textit{multi-loop NCS}, where the centralized scheduler takes a decision based on observed AoI. We re-use the concept of age penalty function, but, unlike the related work, we derive \textit{control-aware age penalty} which take into account control system dynamics. By modeling the system as a Markov decision process (MDP) with AoI as its states, we further derive a $\gamma$-optimal stationary scheduling policy to minimize the network-induced estimation error over an infinite horizon. We show that our scheduling policy outperforms the related work age-optimal scheduling proposed in \cite{hsu2017age} and greedy error scheduling proposed in \cite{vilgelm2017control} at minimizing the average network-induced error.

The remainder of the paper is organized as follows. Section~\ref{sec:systemmodel} presents the system model and formulates the problem. Section~\ref{sec:scheduler} introduces the optimal scheduler design. Section~\ref{sec:evaluation} presents simulative evaluation and benchmarking of the proposed scheduler, and Section~\ref{sec:conclusions} concludes the paper.

\subsection{Notations}
Throughout the paper $\bm{v^T}$ and $\bm{M}^T$ stand for the transpose of a
vector $\bm{v}$ and a matrix $\bm{M}$, respectively. $\tr(.)$ is the trace operator.
The expected value of a random variable $X$ is denoted by $\E[X]$. The euclidean norm of a vector $\bm{v}$ is denoted by $\left\Vert \bm{v} \right\Vert$ with $\left\Vert \bm{v} \right\Vert = \sqrt{\bm{v}^T \bm{v}}$. The normal distribution with mean $\mu$ and standard deviation $\sigma$ is denoted by $\mathcal{N}(\mu, \sigma^2)$. $P(A ~|~ B)$ denotes the occurrence probability of an event $A$ given $B$. $\mathbb{N}_+$ is the set of positive natural numbers.
\section{SCENARIO AND PROBLEM STATEMENT}
\label{sec:systemmodel}
\begin{figure}[t]
	\centering
	\resizebox{.95\columnwidth}{!}{%
		\begin{tikzpicture}[>=latex]
\node (plant) at (0,6) [draw,minimum width=2cm,minimum height=1.5cm, rounded corners=0.1cm, text width=3cm, align=center] {\huge{Plant $\mathcal{P}_i$}};

\node (sensor) at (12,6) [draw,minimum width=2cm,minimum height=1.5cm, rounded corners=0.1cm, text width=3cm, align=center] {\huge{Sensor $\mathcal{S}_i$}};

\node (controller) at (0,2) [draw,minimum width=2cm,minimum height=1.5cm, rounded corners=0.1cm, text width=4cm, align=center] {\huge{Controller $\mathcal{C}_i$}};

\node (estimator) at (6,2) [draw,minimum width=2cm,minimum height=1.5cm, rounded corners=0.1cm, text width=4cm, align=center] {\huge{Estimator $\mathcal{E}_i$}};

\node[cloud, cloud puffs=11, cloud ignores aspect, minimum width=4cm, minimum height=3cm, text width=3cm, align=center, draw] (cloud) at (12, 2) {\huge {Network \\ \vspace*{2mm}} \Large{with scheduler}};

\draw[arrows=-triangle 45,black] (controller.north) -- node[pos=0.5, left] {\huge $u_i[t]$} (plant.south);

\draw[arrows=-triangle 45,black] (plant.east) -- node[pos=0.5, above] {\huge $x_i[t]$} (sensor.west);

\draw[black, dashed] (sensor.south) -- node[] {} (cloud.north);

\draw[arrows=-triangle 45,black, dashed] (cloud.west) -- node[pos=0.5, above] {} (estimator.east);

\draw[arrows=-triangle 45,black] (estimator.west) -- node[pos=0.5, above] {\huge $\hat{x}_i[t]$} (controller.east);

%
%
%
%
\node (s2) at (15.5,2) [draw,minimum width=1cm,minimum height=1cm, rounded corners=0.1cm, text width=1cm, align=center] {\huge{$\mathcal{S}$}};

\node (p2) at (17.5,2) [draw,minimum width=1cm,minimum height=1cm, rounded corners=0.1cm, text width=1cm, align=center] {\huge{$\mathcal{P}$}};

\node (e2) at (15.5,4) [draw,minimum width=1cm,minimum height=1cm, rounded corners=0.1cm, text width=1cm, align=center] {\huge{$\mathcal{E}$}};

\node (c2) at (17.5,4) [draw,minimum width=1cm,minimum height=1cm, rounded corners=0.1cm, text width=1cm, align=center] {\huge{$\mathcal{C}$}};

\draw[arrows=-triangle 45,black] (p2.west) -- node[] {} (s2.east);
\draw[black, dashed] (s2.west) -- node[] {} (14.45,2);
\draw[arrows=-triangle 45,black, dashed] (14,2.7) -- (14, 4) -- node[] {} (e2.west);
\draw[arrows=-triangle 45,black] (e2.east) -- node[] {} (c2.west);
\draw[arrows=-triangle 45,black] (c2.south) -- node[] {} (p2.north);

%
%
%

\end{tikzpicture}
	}
	\caption{Considered scenario with $N$ heterogeneous LTI networked control systems. Solid lines indicate ideal controller-to-plant and plant-to-sensor links. Sensor-to-controller link is closed over a shared wireless channel. Medium access is granted by a scheduler.}
	\label{fig:scenario}
\end{figure}
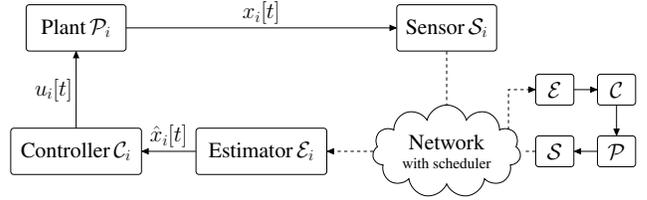

Suppose $\numLoops$ independent, linear time-invariant (LTI) control systems sharing a wireless communication network. Each sub-system $i$ consists of a plant $\plant$, a sensor $\sensor$ and a controller $\controller$ with an estimator $\estimator$. Each sensor $\sensor$ samples the plant periodically and transmits the latest state information to $\controller$. On the controller side, $\estimator$ estimates periodically the current plant state based on the latest received information. Estimated state is then used by $\controller$ to calculate the next control input. We assume ideal links between each controller and plant pair while sensor is operating remotely over the shared wireless channel as illustrated in Fig. \ref{fig:scenario}. 
\subsection{Network Model}
\label{sec:network}
All transmissions are managed by a centralized entity called \textit{scheduler}. In each time slot $t$, the scheduler decides which $R\leq N$ subsystems are allocated transmission resources. Transmission resources are assumed non-overlapping, thus, contention-free medium access without collisions is guaranteed. Let us define a scheduling decision variable $q_i[t] \in \{0, 1\}$ being $q_i[t] = 1$ if the sub-system $i$ is allowed to transmit in slot $t$ and $q_i[t] = 0$ otherwise. Therefore, the following holds for all time slots:
\begin{equation}
	\sum_{i=1}^{N} q_i[t] \leq R \quad ,\forall t.
\end{equation}
 The smallest time unit we consider is a time slot which is also equal to the sampling period of sub-systems. In other words, when a sensor is granted channel access, it always transmits its latest state information. We assume a constant delay of one time slot, i.e., any transmission scheduled in time slot $t$ is assumed to be received in $t+1$. We leverage the results in \cite{costa2014age} and suppose that sensors replace their packet in the transmit buffer with a more recent measurement if they do not receive any resource grant.

We consider \textit{packet erasure channel} where each sub-system $i$ has a constant probability $p_i$ for a successful packet reception and decoding. That is, if we define a success indicator variable $\delta_i[t] \in \{0,1\}$ representing a failed $\delta_i[t] = 0$ or a successful $\delta_i[t] = 1$ reception of a packet, then the success probability is given by $P\left(\delta_i[t]=1 \;\middle|\; q_i[t] = 1\right) = p_i \in (0,1]$. Similarly, packet loss probability is $P\left(\delta_i[t] =0 \;\middle|\; q_i[t] = 1\right) = 1 - p_i$. For the sake of completeness, note that $P\left(\delta_i[t]=1 \;\middle|\; q_i[t] = 0\right) = 0$ and $P\left(\delta_i[t]=0 \;\middle|\; q_i[t] = 0\right) = 1$.

\subsection{Age-of-information Model}
\label{sec:aoi}
\textit{Age-of-information} of a  sub-system $i$ is defined as the time elapsed since the generation of the freshest information available at $\controller$. If we denote AoI of the $i$-th sub-system by $\Delta_i[t]$, it can be obtained from $\Delta_i[t] \triangleq t - r_i$ with $r_i = \sup\{\tau \in \mathbb{N} : \tau < t, \, q_i[t] \cdot \delta_i[t] = 1\}$ with time evolution:
\begin{equation}
	\Delta_i[t+1] =   
	\begin{cases}
		1, & \text{if } q_i[t] \cdot \delta_i[t] = 1\\
		\Delta_i[t] + 1, & \text{otherwise}.
	\end{cases}
	\label{eq:aoi}
\end{equation}
Due to the constant one slot delay, a successfully received packet resets AoI to one, hence $\Delta_i[t] \in \mathbb{N}_+, \, \forall i,t$. Otherwise, AoI increases linearly as time passes. 
 
\subsection{Control System Model}
The system state of the $i$-th plant $\plant$ evolves by the following LTI model over discrete time $t$:
\begin{equation}
	\bm{x}_i[t+1] = \bm{A}_i \bm{x}_i[t] + \bm{B}_i \bm{u}_i[t] + \bm{w}_i[t],
\end{equation}
with state vector $\bm{x}_i \in \mathbb{R}^{n_i}$, control input $\bm{u}_i\in \mathbb{R}^{m_i}$, system and input matrices $\bm{A}_i \in \mathbb{R}^{n_i \times n_i}$ and $\bm{B}_i \in \mathbb{R}^{n_i \times m_i}$, respectively. We assume the identically and independently distributed (i.i.d.) disturbance vector $\bm{w}_i[t] \in \mathbb{R}^{n_i}$ to have zero-mean Gaussian distribution with diagonal covariance matrix $\bm{\Sigma}_i$. Each sub-system is initialized with $\bm{x}_i[0] \sim \mathcal{N}(\bm{0}, \bm{\Sigma}_i)$. 

In the estimator $\estimator$, the current system state is estimated given the latest observed state, i.e., $\bm{\hat{x}}_i[t] \triangleq \E\left[\bm{x}_i[t] \;\middle|\; \bm{x}_i[t - \Delta_i[t]\right]$ as:
\begin{equation}
	\bm{\hat{x}}_i[t] = \bm{A}_i^{\Delta_i[t]} \bm{x}_i\left[t - \Delta_i[t]\right] + \sum_{q = 1}^{\Delta_i[t]} \bm{A}_i^{q-1}\bm{B}_i \bm{u}_i[t-q]
\end{equation}
with $\bm{\hat{x}}_i \in \mathbb{R}^{n_i}$\footnote{The estimator requires the history of control inputs applied to the plant. Since we assume stationary control policy throughout the paper, this assumption does not impose any additional communication between the controller and estimator.}. The control input is obtained from the control law $\bm{u}_i[t] = - \bm{L}_i \bm{\hat{x}}_i[t]$ with feedback gain matrix $\bm{L}_i \in \mathbb{R}^{m_i \times n_i}$\footnote{We assume that the control system is stabilizable and the stabilizing gain $\bm{L}_i$ exists.}. We leverage our results in \cite{ayan2019age} and define state estimation error of $\estimator$ at $t$, i.e., $\bm{e}_i[t] \triangleq \bm{x}_i[t] - \bm{\hat{x}}_i[t]$, as a function of AoI:
\begin{equation}
\bm{e}_i[t]  = \sum_{q=1}^{\Delta_i[t]}\bm{A}_i^{q-1} \bm{w}_i[t-q].
\label{eq:nie}
\end{equation}
Analogously, the mean-square norm of $\bm{e}_i[t]$ is derived as:
\begin{equation}
	\E \left[ \left\Vert \bm{e}_i[t] \right\Vert^2\right] = \sum_{r=0}^{\Delta_i[t] - 1} \tr\left( (\bm{A}_i^T)^r \bm{A}_i^r \bm{\Sigma_i} \right) \triangleq g_i(\Delta_i[t]).
	\label{eq:mse}
\end{equation}
It is important to note that $\Delta_i[t]$ is the only time dependent variable in \eqref{eq:mse} since we deal with time-invariant control systems. In addition, $\E \left[ \left\Vert \bm{e}_i[t] \right\Vert^2\right]$ is an increasing function in age. Therefore, by providing the knowledge of the current state to $\estimator$ (or any fresher information than the one it already has), the reduction in $\E \left[ \left\Vert \bm{e}_i[t] \right\Vert^2\right]$ is guaranteed.


\subsection{Scheduling Problem Formulation}
We model the centralized scheduling problem for $N$ networked control systems as a \textit{markov decision process (MDP)}~\cite{bertsekas2017dynamic}. We define the vector of sub-systems' AoIs as a state of the MDP, i.e., $\bm{s}[t] \triangleq (\Delta_1[t], \, \Delta_2[t], \, \dots \,, \, \Delta_N[t])$. Note that the state space $\statespace$ is equal to $\mathbb{N}_+^N$ thus by definition is a countable $N$-dimensional infinite set. The action space is defined as the set of all possible scheduling decisions of length $N$, i.e., $\bm{\mathcal{A}'} = \{0, \, 1\}^N$. Here, the $i$-th index of $\bm{a}[t]$ being 1 indicates that the $i$-th sub-system is scheduled for transmission, i.e., $a_i[t] = q_i[t], \forall i \in \mathbb{N}^+$. Note that, since the scheduler has to select $R$ out of $N$ sub-systems to transmit, we define the set of admissible actions as $\bm{\mathcal{A}} = \{ \bm{a} \in \bm{\mathcal{A}'} : \sum_{i=1}^{N} a_i \leq R\}$. 

%
%

Given the states and admissible actions as above, we are interested in deterministic scheduling policies $\bm{\pi}$ within the set of admissible policies $\bm{\Pi}$ that maps states into admissible actions, i.e., $\bm{\Pi}: \bm{\mathcal{S}} \mapsto \bm{\mathcal{A}}$. Our goal is to find an optimal policy $\bm{\pi}^* \in \bm{\Pi}$ that minimizes the \textit{total mean-square estimation error over an infinite horizon}, given any initial state $\bm{s}_0 = (\Delta_1[0], \, \dots, \, \Delta_N[0])$:
\begin{equation}
J_{\bm{\pi}^*} (\bm{s}_0) = \min_{\bm{\pi} \in \bm{\Pi}} ~ J_{\bm{\pi}}(\bm{s}_0)
\label{eq:optProb}
\end{equation}
with
\begin{equation}
	J_{\bm{\pi}}(\bm{s}_0) = \lim_{T \rightarrow \infty} \E_\pi \left[ \sum_{t=0}^{T-1} \sum_{i=1}^{N}  \gamma^t g_i(\Delta_i[t])\right],
	\label{eq:discountedCostProblem}
\end{equation}
where $\E_{\bm{\pi}}$ indicates the expected mean-square error under policy $\bm{\pi}$. Here, $\gamma \in (0, 1)$ is a positive scalar value called $\textit{discount factor}$.
\begin{mydef}
	A scheduling policy is called $\gamma$-optimal if it minimizes the discounted mean-square estimation error over an infinite horizon, i.e., the right hand side of \eqref{eq:discountedCostProblem}~\cite{bertsekas2017dynamic}.
\end{mydef}
The role of $\gamma$ is to adjust the importance of short-term or long-term penalties which are the total mean-square error norms as in \eqref{eq:mse}. A penalty received in $t$ steps in the future is worth only $\gamma^t$ times what it would cost if it were received immediately. Hence, as $\gamma$ approaches 1, the scheduler weights future penalties more and thereby becomes more farsighted.
\section{SCHEDULER DESIGN}
\label{sec:scheduler}

\subsection{Approximating Sequence of the MDP}
\label{sec:approximation}
The state space of the given MDP is infinite. Since we can not iterate through infinite states in practice, we work with an \textit{augmentation type approximating sequence} (ATAS) of the MDP \cite{sennott1999stochastic}. First, let us define a nonempty finite state space $\approxstatespace \subset \statespace$ that divides $\statespace$ into two disjoint sets, where states with $\Delta_i \leq M , \, \forall i$ belong to the finite set $\approxstatespace$ and states with $\Delta_i > M, \, \exists i$ belong to $\statespace\setminus\bm{\mathcal{S}}_M$. In other words, as long as AoIs of all sub-systems are smaller than $M + 1$, the state is in $\approxstatespace$. If the AoI of at least one sub-system is larger than $M$, the state is in the infinite set $\statespace\setminus\bm{\mathcal{S}}_M$.

Suppose that in state $\bm{s} \in \approxstatespace$ action $\bm{a}$ is chosen. For any $\bm{s'} \in \approxstatespace$ the transition probability from $\bm{s}$ to $\bm{s'}$ $P_{\bm{ss'}}(\bm{a})$ is unchanged. Now let us define the \textit{excess probability} as the transition probability from a state $\bm{s} \in \approxstatespace$ to a state $\hat{\bm{s}} \notin \approxstatespace$ outside the finite space with $P_{\bm{s}\hat{\bm{s}}}(a) \geq 0$. In this case we redistribute all excess probabilities to the states $\bm{s'}$ in $\bm{S}_M$ according to a probability distribution $f_{\bm{s'}}(\bm{s},\bm{a}, \hat{\bm{s}}, M)$. If we denote the resulting MDP by MDP$_M$ the definition of ATAS follows from~\cite{sennott1999stochastic}.

\begin{mydef}
	A sequence MDP$_M$ is an augmentation type approximating sequence of an MDP if for each $\hat{\bm{s}} \notin \approxstatespace$, given $\bm{s}, \bm{s'} \in \approxstatespace$ and $\bm{a} \in \bm{\mathcal{A}}$, there exists a probability distribution $f_{\bm{s'}}(\bm{s}, \bm{a}, \hat{\bm{s}}, M)$ such that
	\begin{equation}
		P_{\bm{ss'}}(\bm{a};M) = P_{\bm{ss'}}(\bm{a}) + \sum_{\hat{\bm{s}} \in \statespace\setminus\bm{\mathcal{S}}_M} P_{\bm{s}\hat{\bm{s}}}(\bm{a})f_{\bm{s'}}(\bm{s}, \bm{a}, \hat{\bm{s}}, M).
	\end{equation}
\end{mydef}

\begin{figure}[t]
	\centering
	\resizebox{.7\columnwidth}{!}{%
		\begin{tikzpicture}[>=latex]
\node (s0) at (0,0) [draw, circle, align=center, inner sep=0pt, minimum size=1cm] {$1$};
\node (sM) at (2,0) [draw, circle, align=center, inner sep=0pt, minimum size=1cm] {$M$};
\node (sM1) at (4,0) [draw, circle, align=center, inner sep=0pt, minimum size=1cm] {$M+1$};

\node () at (1,0) [draw, fill=black, circle, align=center, inner sep=0pt, minimum size=0.1cm] {};
\node () at (1.3,0) [draw, fill=black, circle, align=center, inner sep=0pt, minimum size=0.1cm] {};
\node () at (.7,0) [draw, fill=black, circle, align=center, inner sep=0pt, minimum size=0.1cm] {};

\draw[->] (sM) [out=40,in=140, dashed] to node [pos=0.1, near end, above] (aa) {$0$} (sM1);
\draw[->] (sM) [out=40,in=100,loop] to node [pos=0.1, midway, above] (aa) {$1-p_1$} (M);
\draw[->] (sM.south) [out=-140,in=-40] to node [midway, below] (m2) {$p_1$} (s0.south);
\end{tikzpicture}
	}
	\caption{Augmentation type approximating sequence for the exemplary MDP with one syb-system $i=1$ ($N=1$) with maximum augmented age $M$. Given the current state $\bm{s}[t] = M$ and action $\bm{a}[t]$ with $a_1[t] = q_1[t] = 1$, we redistribute the excess probability of a transition from $M$ to $M+1$ back to state $M$, i.e., $P\left(\bm{s}[t+1] = M \; \middle| \; \bm{s}[t] = M, a_1[t] = 1\right) = 1 - p_1$.}
	\label{fig:approximateMDP}
\end{figure}
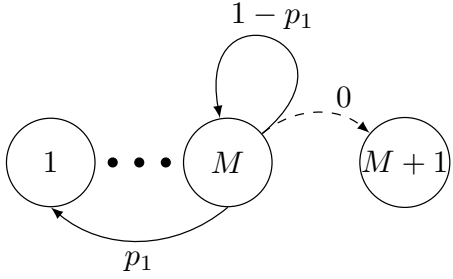
By this definition we introduce the following ATAS for our scenario: Let $\tilde{\Delta}_i^{(M)}[t]$ be the \textit{augmented age-of-information} for the $i$-th sub-system with $\tilde{\Delta}^{(M)}_i[t] = \min(\Delta_i[t], \, M)$. That is, we truncated AoI to $M$ if its value exceeds $M$. Thus the new state space becomes finite with $\approxstatespace = \{1, \, \dots, \, M\}^N$. Fig. \ref{fig:approximateMDP} illustrates the resulting MDP$_{M}$ for a single sub-system $i$ given the scheduler decides to schedule the same user at age $M$, i.e., $a_i[t] = 1$. In the remainder of this paper we refer to the augmented states without changing the notation, $\bm{s} \in \approxstatespace $ to avoid visual clutter.

As we have the states, actions, and transition probabilities of the finite MDP$_{M}$, let us define the immediate cost associated with each augmented state and action pair as:
\begin{equation}
\label{eq:apprxCost}
	C(\bm{s}[t], \bm{a}[t]) = \sum_{i=1}^{N} g_i(\tilde{\Delta}^{(M)}_i[t]), 
\end{equation}
with states $\bm{s} \in \approxstatespace$. Note that cost does not depend on the action taken. One can easily introduce an additive communication cost term to the equation above in case different sub-systems demand different amount of network resources. 
 \subsection{$\gamma$-Optimal Discounted Error Scheduler (DES)}
 \label{subsec:discountedscheduler}
 Now for the given augmentation level $M$, we propose a $\gamma$-optimal stationary deterministic scheduling policy that minimizes the discounted infinite horizon problem from \eqref{eq:optProb} with cost approximations from \eqref{eq:apprxCost}\footnote{Note that, the proposed algorithm solves the approximated problem optimally for a given $\gamma$ and an augmentation level $M$.}. To that end, we apply the standard \textit{value iteration} approach with \textit{dynamic programming} \cite{bertsekas2017dynamic}. 
 
 Each state is associated with an initial value function $J_0(\bm{s})$. We iterate through all state-action pairs $(\bm{s}, \bm{a})$ and their possible next states $\bm{s}' $ as:
 \begin{align}
 \label{eq:Bellman}
 	J_{k+1}(\bm{s}) &= \min_{\bm{a} \in \bm{\mathcal{A}}} \left\{ \E\left[C(\bm{s}, \bm{a}) + \gamma  J_k(\bm{s}') \right] \right\} \nonumber \\
 				    &= \min_{\bm{a} \in \bm{\mathcal{A}}} \left\{C(\bm{s}, \bm{a}) + \gamma  \E \left[ J_k(\bm{s}') \right]\right\} \nonumber \\
 				    &= \min_{\bm{a} \in \bm{\mathcal{A}}} \left\{C(s, \bm{a}) + \gamma \sum_{\bm{s}' \in \approxstatespace}  P_{ss'}(\bm{a}) J(\bm{s}') \right\},
 \end{align}
while minimizing the right-hand-side of the Bellman equation \eqref{eq:Bellman}. Here, $P_{ss'}(\bm{a})$ denotes the success probability from a current state $\bm{s}$ to a successor state $\bm{s}'$ given action $\bm{a}$, i.e., $P\left(\bm{s}[t+1] =  \bm{s}' \; \middle| \; \bm{s}[t] = \bm{s}, \, \bm{a}[t] = \bm{a}\right)$. In \cite{bertsekas2017dynamic} it is shown that, as $k$ goes to infinity, value functions will converge to an optimal cost, i.e., $\lim_{k \rightarrow \infty} J_k =  J^*$, for any initial value function $J_0$. In this case actions solving \eqref{eq:Bellman} constitute the stationary optimal policy $\pi^*$. Algorithm \ref{alg:valueiteration} illustrates the practical implementation of the value iteration algorithm with $J_0(\bm{s}) = 0, ~ \forall \bm{s}$. A convergence threshold $\theta$ is used to break the loop. The precision of the final value functions increases if lower values of $\theta$ are selected.

\begin{algorithm}[t]
$J(\bm{s}) \leftarrow 0$ for all augmented states $\bm{s} \in \approxstatespace$

	\Do{$u > \theta$}{
		
		$u \leftarrow 0$
		
		\ForEach{$\bm{s} \in \approxstatespace$}{
			Update $J^+(\bm{s})$ as in \eqref{eq:Bellman}\;
			
			$u \leftarrow \max(u, \, |J^+(\bm{s}) - J(\bm{s})|)$ \;
			$J(\bm{s}) \leftarrow J^+(\bm{s})$\;
		}
		
	}
	$\pi^*(\bm{s}) \leftarrow \underset{\bm{a} \in \mathcal{A}}{\arg \min} ~ J(\bm{s})$
	
	\KwOut{$\pi^* $}
	
	\caption{Value iteration}
	\label{alg:valueiteration}
\end{algorithm}

\section{EVALUATION}
\label{sec:evaluation}
In this section, we present a simulative analysis and compare our proposed scheduler to selected benchmark schedulers from the existing literature. Section \ref{subsec:simulation} describes the simulation setup and parameters. Sections \ref{subsec:benchmarks} and \ref{subsec:results} introduce the benchmark schedulers and discuss simulation results.

\subsection{Benchmarks}
\label{subsec:benchmarks}
In order to compare the performance of our proposed scheduler, we choose two representative schedulers from the state-of-the-art: the $\gamma$-optimal AoI scheduler proposed in~\cite{hsu2017age} and the control-aware greedy scheduler proposed in~\cite{vilgelm2017control}.

\subsubsection{$\gamma$-Optimal AoI Scheduler (AoIS)}
\label{subsec:aoischeduler}
As the name suggests, the \textit{$\gamma$-optimal AoI scheduler} employs the total age-of-information in the network as the immediate cost of a state:
\begin{equation}
\label{eq:AoICost}
C_{AoI}(\bm{s}[t], \bm{a}[t]) = \sum_{i=1}^{N}\tilde{\Delta}^{(M)}_i[t].
\end{equation}
The $\gamma$-optimal AoI policy can be obtained by simply replacing the cost function in Alg. \ref{alg:valueiteration} with the new one from \eqref{eq:AoICost}. As we are going to show in Section \ref{subsec:results}, the resulting policy provides the highest information freshness among the selected schedulers.

\subsubsection{Greedy Error Scheduler (GES)}
\label{subsec:greedyscheduler}
\begin{figure}
	\centering
	\begin{subfigure}[t]{0.3\columnwidth}
		\includegraphics[width=\columnwidth]{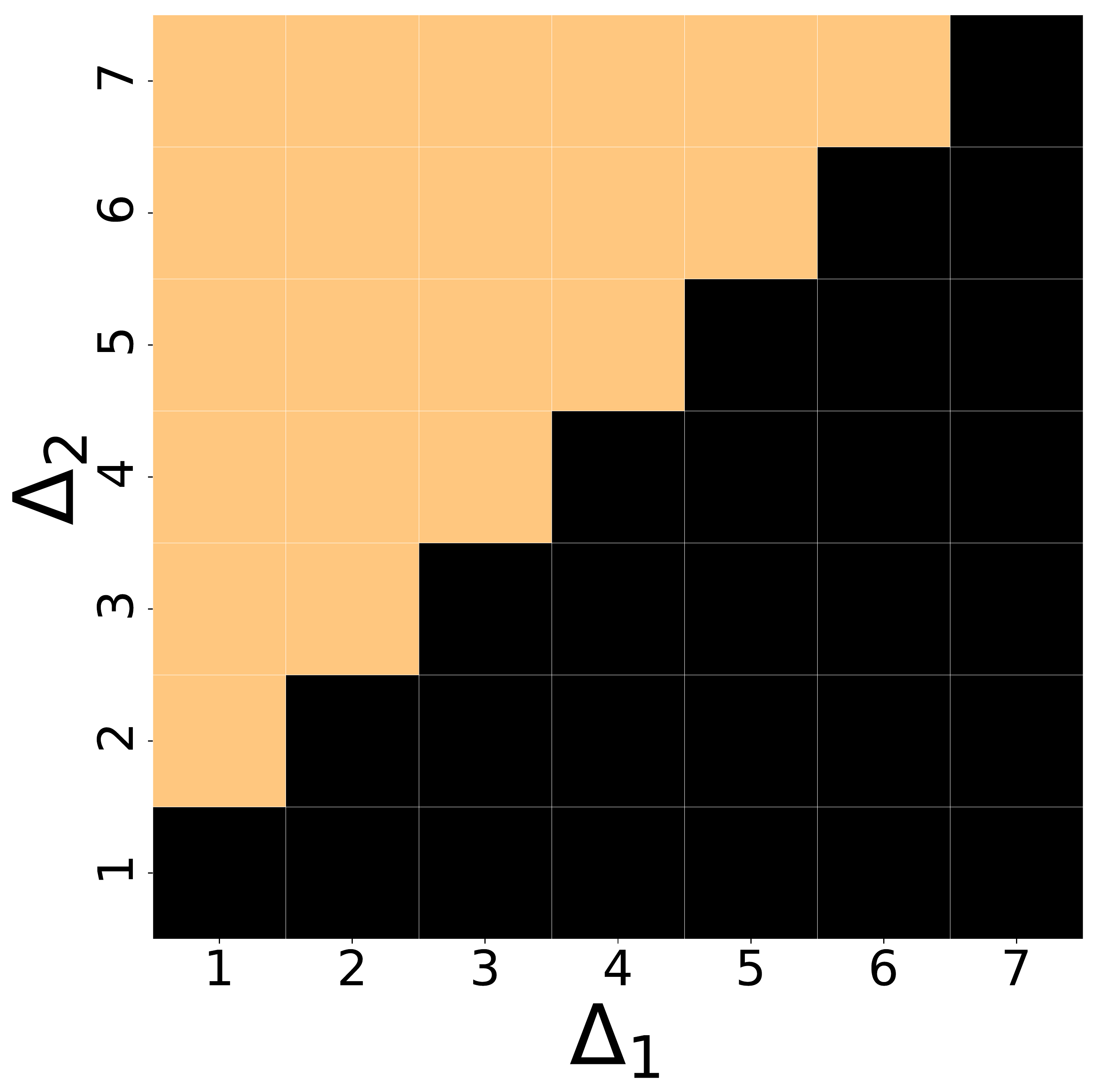}
		\caption{AoIS policy}
		\label{fig:AoIpolicy}
	\end{subfigure}
	~ 
	\begin{subfigure}[t]{0.3\columnwidth}
		\includegraphics[width=\columnwidth]{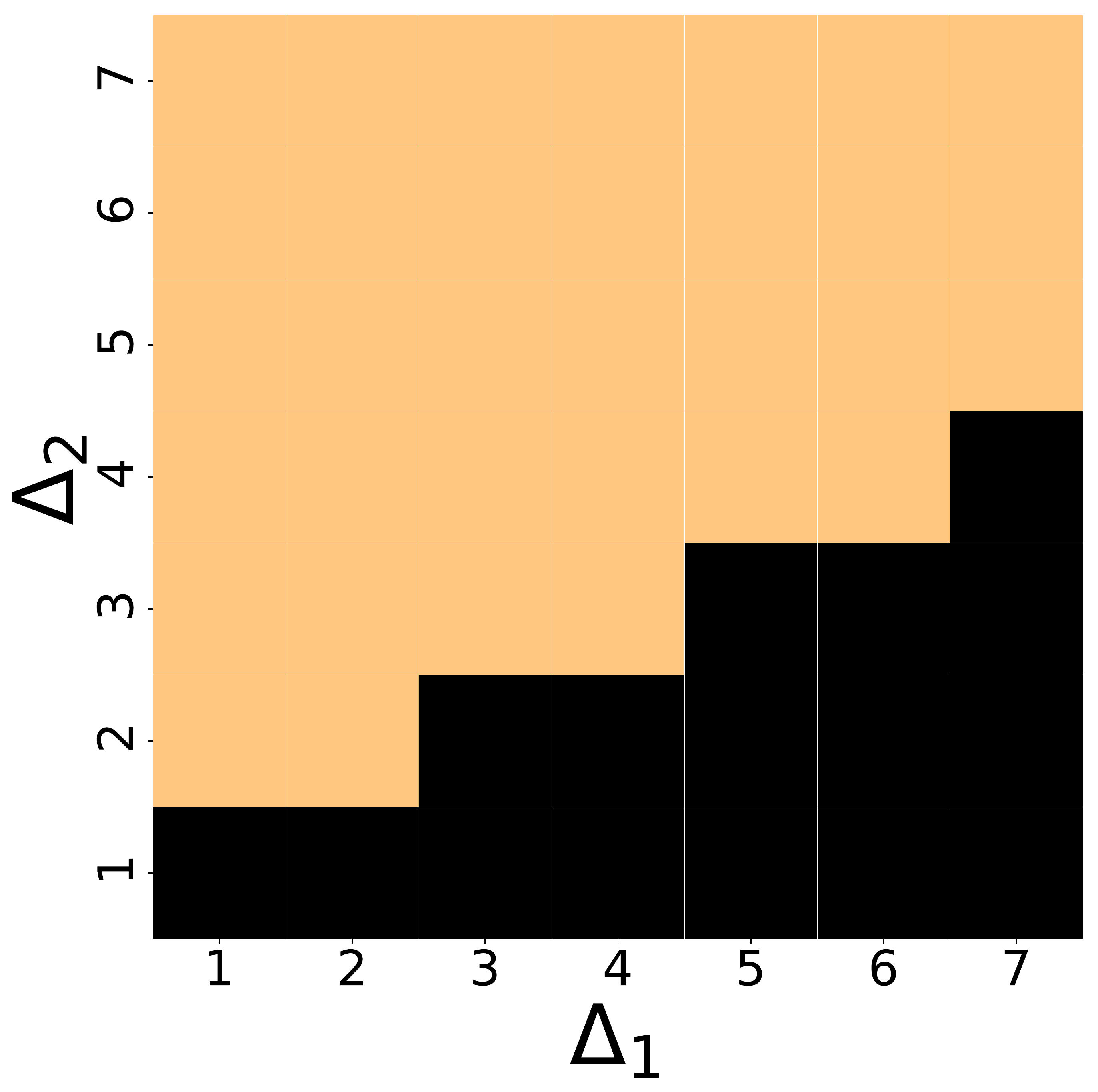}
		\caption{GES policy}
		\label{fig:greedyVoIpolicy}
	\end{subfigure}
	~ 
	\begin{subfigure}[t]{0.3\columnwidth}
		\includegraphics[width=\columnwidth]{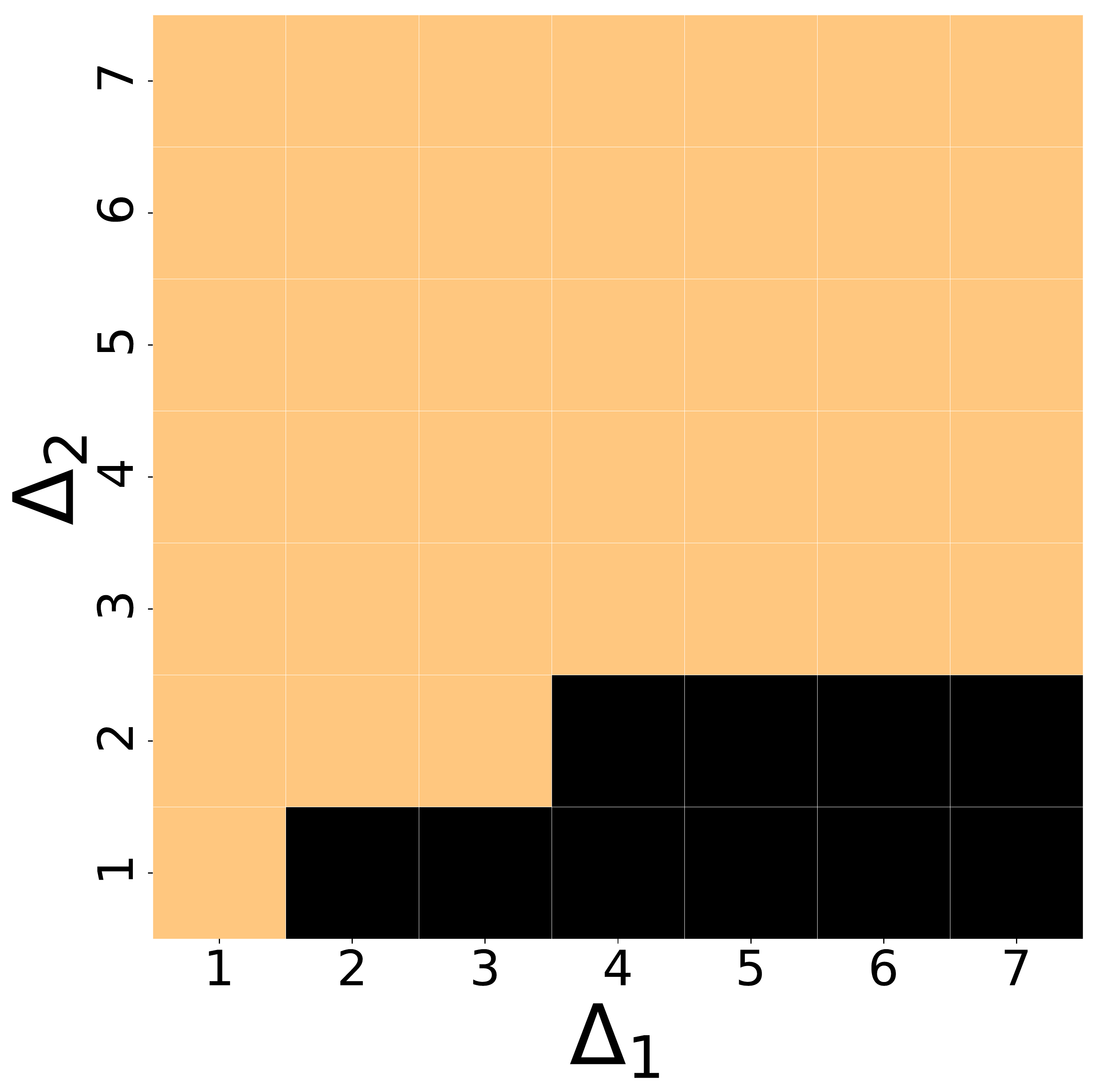}
		\caption{DES policy}
		\label{fig:VoI_policy}
	\end{subfigure}
	\caption{Scheduling policies of (a) AoIS, (b) GES, and (c) DES for two sub-systems with state matrices $A_1 = 1.1$ and $A_2 = 1.3$, loss probabilities $p_1 = p_2 = 0.5$, and with a single resource per time slot $R=1$. Parameters: $\gamma = 0.9$, $M=7$, $\theta=0.1$. Light squares correspond to states $\bm{s}=\left(\Delta_1,\Delta_2\right)$ where user 1 is scheduled (${a}_1=1,{a}_2=0$), and dark squares correspond to states where user 2 is scheduled (${a}_1=0,{a}_2=1$).}
	\label{fig:policies}
\end{figure}

The \textit{greedy error scheduler} prioritizes sub-systems with the highest mean squared error:
\begin{equation}
\bm{a}[t] = \arg \max_{\bm{a} \in \bm{\mathcal{A}}} \left\{ \sum_{i}^{N}  p_i  g_i(\Delta_i[t]) \right\}.
\end{equation}
Here, we introduce the success probabilities as a weighting factor for a \textit{channel-aware} version of the proposed solution in \cite{vilgelm2017control}. One can simply ignore it for a \textit{control-aware} implementation only. Fig. \ref{fig:policies} illustrates optimal policies of AoIS, GES, and DES for a simplified scenario with $N=2$ heterogeneous control loops and one resource $R=1$. Fig. \ref{fig:AoIpolicy} is symmetrical since AoIS is independent of control system parameters. Fig. \ref{fig:greedyVoIpolicy} and Fig. \ref{fig:VoI_policy} highlight the asymmetrical distribution of resources among both control loops coming from their discrepancy in system instability.

\subsection{Simulation Set-Up}
\label{subsec:simulation}
Suppose the system comprises $N=5$ heterogeneous control sub-systems sharing a wireless channel with Bernoulli-distributed random packet loss. We assume that all control sub-systems have different plant system dynamics, with increasingly unstable system matrices. To provide intuitive and illustrative results, we consider scalar sub-systems in the evaluation. To that end, we select $A_i = \{1.1, \, 1.3, \, 1.5, \, 1.7, \, 1.9\}$ for $i=\{1,\, 2,\, 3,\, 4,\, 5\}$, respectively. We assume equal packet success probability for each sub-system $i$, i.e., $p_i = 0.9, \; \forall i$. The feedback gain matrix is given by $L_i = A_i$ which corresponds to the deadbeat control strategy. Input and noise covariance matrices are selected as $B_i = 1$ and $\Sigma_i = 1$, respectively. We further consider single transmission resource per time slot, i.e., $R = 1$. The augmentation level for the approximating sequence of the original MDP is $M = 25$, and the stop condition for Alg.~\ref{alg:valueiteration} is $\theta = 0.1$. Simulations are run for $T=20000$ time slots. The discount factor is varied, $\gamma = \{0.1, \, 0.2, \dots, 0.9\}$ to study its effect on the resulting average quadratic error per time slot, $\overline{e}$, and average age per time slot, $\overline{\Delta}$, given by:
\begin{align}
\overline{e} &=
\dfrac{1}{T} \dfrac{1}{N} \sum_{t = 0}^{T-1}\sum_{i=1}^{N} \lVert e_i[t]\rVert^2 \\
\overline{\Delta} &=
\dfrac{1}{T} \dfrac{1}{N} \sum_{t = 0}^{T-1}\sum_{i=1}^{N} \Delta_i[t]
\end{align}

\subsection{Results}
\label{subsec:results}

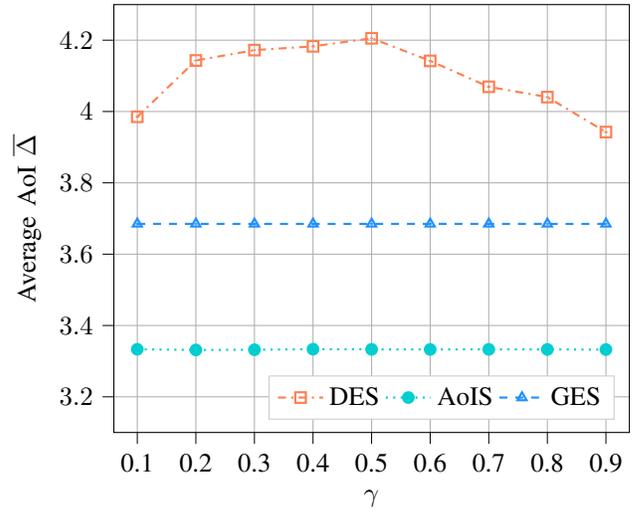
\begin{figure}[t]
	\centering
\begin{tikzpicture}

\definecolor{color0}{rgb}{1,0.498039215686275,0.313725490196078}
\definecolor{color1}{rgb}{0,0.807843137254902,0.819607843137255}
\definecolor{color2}{rgb}{0.117647058823529,0.564705882352941,1}

\begin{axis}[
legend cell align={left},
legend style={at={(0.3,0.03)}, anchor=south west, draw=white!80.0!black},
legend columns=3,
tick align=outside,
tick pos=left,
x grid style={white!69.01960784313725!black},
xlabel={\(\displaystyle \gamma\)},
xmajorgrids,
xmin=0.06, xmax=0.94,
xtick style={color=black},
xtick={0,0.1,0.2,0.3,0.4,0.5,0.6,0.7,0.8,0.9,1},
xticklabels={0.0,0.1,0.2,0.3,0.4,0.5,0.6,0.7,0.8,0.9,1.0},
y grid style={white!69.01960784313725!black},
ylabel={Average AoI $\overline{\Delta}$},
ymajorgrids,
ymin=3.1, ymax=4.3,
ytick style={color=black}
]
\path [draw=color0, very thick]
(axis cs:0.1,3.9824687847146)
--(axis cs:0.1,3.9879472152854);

\path [draw=color0, very thick]
(axis cs:0.2,4.14112964688641)
--(axis cs:0.2,4.14498295311359);

\path [draw=color0, very thick]
(axis cs:0.3,4.17025940285624)
--(axis cs:0.3,4.17408919714376);

\path [draw=color0, very thick]
(axis cs:0.4,4.18062114496135)
--(axis cs:0.4,4.18482425503865);

\path [draw=color0, very thick]
(axis cs:0.5,4.20358468771976)
--(axis cs:0.5,4.20710871228024);

\path [draw=color0, very thick]
(axis cs:0.6,4.13961723520966)
--(axis cs:0.6,4.14429076479034);

\path [draw=color0, very thick]
(axis cs:0.7,4.06686332251277)
--(axis cs:0.7,4.07178247748722);

\path [draw=color0, very thick]
(axis cs:0.8,4.03827725424373)
--(axis cs:0.8,4.04255934575627);

\path [draw=color0, very thick]
(axis cs:0.9,3.94007384732698)
--(axis cs:0.9,3.94474235267302);

\path [draw=color1, very thick]
(axis cs:0.1,3.33163340493232)
--(axis cs:0.1,3.33493739506768);

\path [draw=color1, very thick]
(axis cs:0.2,3.32979894708072)
--(axis cs:0.2,3.33282985291928);

\path [draw=color1, very thick]
(axis cs:0.3,3.33052899185725)
--(axis cs:0.3,3.33350480814274);

\path [draw=color1, very thick]
(axis cs:0.4,3.33196539367134)
--(axis cs:0.4,3.33501400632866);

\path [draw=color1, very thick]
(axis cs:0.5,3.33154879392075)
--(axis cs:0.5,3.33449480607925);

\path [draw=color1, very thick]
(axis cs:0.6,3.33118738989711)
--(axis cs:0.6,3.33433381010289);

\path [draw=color1, very thick]
(axis cs:0.7,3.33143002189878)
--(axis cs:0.7,3.33462977810122);

\path [draw=color1, very thick]
(axis cs:0.8,3.33118995566832)
--(axis cs:0.8,3.33445484433168);

\path [draw=color1, very thick]
(axis cs:0.9,3.33087485625975)
--(axis cs:0.9,3.33400894374025);

\path [draw=color2, very thick]
(axis cs:0.1,3.68272221231332)
--(axis cs:0.1,3.68709098768668);

\path [draw=color2, very thick]
(axis cs:0.2,3.68272221231332)
--(axis cs:0.2,3.68709098768668);

\path [draw=color2, very thick]
(axis cs:0.3,3.68272221231332)
--(axis cs:0.3,3.68709098768668);

\path [draw=color2, very thick]
(axis cs:0.4,3.68272221231332)
--(axis cs:0.4,3.68709098768668);

\path [draw=color2, very thick]
(axis cs:0.5,3.68272221231332)
--(axis cs:0.5,3.68709098768668);

\path [draw=color2, very thick]
(axis cs:0.6,3.68272221231332)
--(axis cs:0.6,3.68709098768668);

\path [draw=color2, very thick]
(axis cs:0.7,3.68272221231332)
--(axis cs:0.7,3.68709098768668);

\path [draw=color2, very thick]
(axis cs:0.8,3.68272221231332)
--(axis cs:0.8,3.68709098768668);

\path [draw=color2, very thick]
(axis cs:0.9,3.68272221231332)
--(axis cs:0.9,3.68709098768668);

\addplot [thick, color0, dashdotted, mark=square, mark size=2, mark options={solid,fill=none}]
table {%
0.1 3.985208
0.2 4.1430563
0.3 4.1721743
0.4 4.1827227
0.5 4.2053467
0.6 4.141954
0.7 4.0693229
0.8 4.0404183
0.9 3.9424081
};
\addlegendentry{DES}
\addplot [thick, color1, dotted, mark=*, mark size=2, mark options={solid,fill=none}]
table {%
0.1 3.3332854
0.2 3.3313144
0.3 3.3320169
0.4 3.3334897
0.5 3.3330218
0.6 3.3327606
0.7 3.3330299
0.8 3.3328224
0.9 3.3324419
};
\addlegendentry{AoIS}
\addplot [thick, color2, dashed, mark=triangle, mark size=2, mark options={solid,rotate=0,fill=none}]
table {%
0.1 3.6849066
0.2 3.6849066
0.3 3.6849066
0.4 3.6849066
0.5 3.6849066
0.6 3.6849066
0.7 3.6849066
0.8 3.6849066
0.9 3.6849066
};
\addlegendentry{GES}
\end{axis}

\end{tikzpicture}
	\caption{Achieved average age-of-information, i.e., $\overline{\Delta}$, vs. discount factor $\gamma$ for discounted error (DES), $\gamma$-optimal AoI (AoIS) and greedy error (GES) schedulers. Out of five control loops, only one sub-system is allowed to transmit simultaneously due to single channel assumption, i.e., $N=5$ and $R=1$. Success probability of each transmission is $p_i = 0.9 ~ \forall i$. 95\% confidence intervals are too small to be plotted.}
	\label{fig:avgAoI}
\end{figure}
\begin{figure}[t]
	\centering
	\input{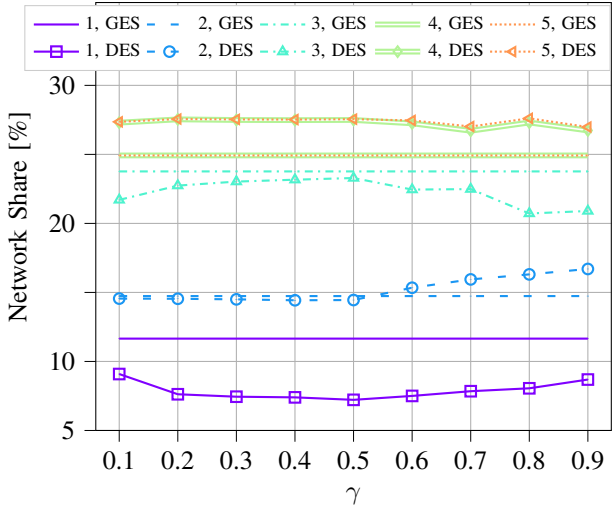}
	\caption{Illustration of network resources shares for different sub-system classes over 100 simulation runs. Numbering in the legend indicate the class $i$ with the respective system matrices $A_{1,2,3,4,5} = \{1.1, 1.3, 1.5, 1.7, 1.9\}$. Lines with and without markers belonging to the same class $i$ illustrate the average network resource share granted to $i$ by the discounted (DES) and greedy (GES) error schedulers, respectively. 95\% confidence intervals are too small to be displayed.}
	\label{fig:networkShares}
\end{figure}

We evaluate the performance of the proposed scheduler using average quadratic error, i.e., $\overline{e}$, as the main metric quantifying estimation accuracy and average age-of-information, i.e., $\overline{\Delta}$, measuring the information freshness\footnote{Results were generated via Monte Carlo experiments with 100 repetitions for any given $\gamma$. Each experiment consists of $T=20000$ time slots.}. In Fig.~\ref{fig:avgAoI}, we plot the average age-of-information $\overline{\Delta}$ against increasing values of the discount factor $\gamma$, corresponding to the decreasing discount of future state. As expected, we observe that the AoI scheduler (AoIS) outperforms the greedy error scheduler (GES) and the $\gamma$-optimal discounted error scheduler (DES) with respect to information freshness at the receiver nodes. An interesting finding is that DES results in a concave shape with respect to $\overline{\Delta}$ as $\gamma$ increases. This can be elaborated with the help of Fig. \ref{fig:networkShares} which illustrates average network shares of individual sub-systems with different task-criticalities. A network share $\alpha_i \%$ indicates that the $i$-th sub-system was granted channel access $\alpha_i \%$ of the time in average. In the figure, we plot  network shares of individual classes $i \in \{1, \dots, 5\}$ with system matrices $A_i = \{1.1, 1.3, 1.5, 1.7, 1.9\}$ against varying $\gamma$. One can observe that the discount factor has different impact on different classes. Increasing $\gamma$ from $0.1$ to $0.5$ has no impact on classes two, four, and five, but it prioritizes class 3 over class 1 (hence, its value is increasing). Therefore, the gap between the sub-systems with $A_1 = 1.1$ and $A_3 = 1.5$ grows. This leads to reduced fairness with respect to resource distribution and thus the growth in AoI follows for the DES. Further increase of $\gamma$, however, reverses this behavior and, at the same time, increases the priority and network share of class 2. Thus the resource distribution among all classes get closer and the decrease in AoI follows. On the other hand, we see that the GES allocates resources in a relatively fairer fashion compared to the DES. This results in a lower average AoI for GES in Fig. \ref{fig:avgAoI}. In addition, the control system unawareness of AoIS leads to equal distribution of resources among different classes. In other words, network shares of all five loops coincide at 20\%. However, we ignore the corresponding figure due to space considerations.

\begin{figure}[t]
	\centering
\begin{tikzpicture}

\definecolor{color0}{rgb}{1,0.498039215686275,0.313725490196078}
\definecolor{color1}{rgb}{0,0.807843137254902,0.819607843137255}
\definecolor{color2}{rgb}{0.117647058823529,0.564705882352941,1}

\begin{axis}[
legend cell align={left},
legend style={at={(0.97,0.5)}, anchor=east, draw=white!80.0!black, font=\small},
log basis y={10},
tick align=outside,
tick pos=left,
x grid style={white!69.01960784313725!black},
xlabel={\(\displaystyle \gamma\)},
xmajorgrids,
xmin=0.06, xmax=0.94,
xtick style={color=black},
xtick={0,0.1,0.2,0.3,0.4,0.5,0.6,0.7,0.8,0.9,1},
xticklabels={0.0,0.1,0.2,0.3,0.4,0.5,0.6,0.7,0.8,0.9,1.0},
y grid style={white!69.01960784313725!black},
ylabel={Average quadratic error \(\displaystyle \overline{e}\)},
ymajorgrids,
yminorticks=true,
ymode=log,
ymin=25.8815885574753, ymax=110,
ytick style={color=black},
ytick={30,40,50, 60, 70, 80, 90, 100},
yticklabels={\(\displaystyle {30}\),\(\displaystyle {40}\),\(\displaystyle {}\), \(\displaystyle {60}\), \(\displaystyle {}\), \(\displaystyle {80}\), \(\displaystyle {}\), \(\displaystyle {100}\)}
]
\path [draw=color0, very thick]
(axis cs:0.1,30.684359974467)
--(axis cs:0.1,31.6658414181689);

\path [draw=color0, very thick]
(axis cs:0.2,30.9019018311362)
--(axis cs:0.2,31.3046816256339);

\path [draw=color0, very thick]
(axis cs:0.3,30.9478809231564)
--(axis cs:0.3,31.4297739121967);

\path [draw=color0, very thick]
(axis cs:0.4,30.8430150555296)
--(axis cs:0.4,31.3114130731172);

\path [draw=color0, very thick]
(axis cs:0.5,30.872807934486)
--(axis cs:0.5,31.4368287303864);

\path [draw=color0, very thick]
(axis cs:0.6,30.3686363748133)
--(axis cs:0.6,30.9483075499822);

\path [draw=color0, very thick]
(axis cs:0.7,29.4882093235807)
--(axis cs:0.7,30.3059004678963);

\path [draw=color0, very thick]
(axis cs:0.8,28.6771790556013)
--(axis cs:0.8,29.0517791023127);

\path [draw=color0, very thick]
(axis cs:0.9,27.5314772527112)
--(axis cs:0.9,28.1307486462022);

\path [draw=color1, very thick]
(axis cs:0.1,86.8368955848851)
--(axis cs:0.1,94.3060467920788);

\path [draw=color1, very thick]
(axis cs:0.2,86.0545345970025)
--(axis cs:0.2,92.9754980359574);

\path [draw=color1, very thick]
(axis cs:0.3,85.034897389385)
--(axis cs:0.3,97.0121274649082);

\path [draw=color1, very thick]
(axis cs:0.4,87.4774387330701)
--(axis cs:0.4,101.721041872776);

\path [draw=color1, very thick]
(axis cs:0.5,81.3814734605198)
--(axis cs:0.5,109.470026235859);

\path [draw=color1, very thick]
(axis cs:0.6,86.1880659756613)
--(axis cs:0.6,92.544515703008);

\path [draw=color1, very thick]
(axis cs:0.7,85.1571270601773)
--(axis cs:0.7,105.145667202271);

\path [draw=color1, very thick]
(axis cs:0.8,86.8338908364612)
--(axis cs:0.8,93.1826835373445);

\path [draw=color1, very thick]
(axis cs:0.9,87.353281059816)
--(axis cs:0.9,91.5441661108329);

\path [draw=color2, very thick]
(axis cs:0.1,32.3741425176901)
--(axis cs:0.1,33.024682540071);

\path [draw=color2, very thick]
(axis cs:0.2,32.3741425176901)
--(axis cs:0.2,33.024682540071);

\path [draw=color2, very thick]
(axis cs:0.3,32.3741425176901)
--(axis cs:0.3,33.024682540071);

\path [draw=color2, very thick]
(axis cs:0.4,32.3741425176901)
--(axis cs:0.4,33.024682540071);

\path [draw=color2, very thick]
(axis cs:0.5,32.3741425176901)
--(axis cs:0.5,33.024682540071);

\path [draw=color2, very thick]
(axis cs:0.6,32.3741425176901)
--(axis cs:0.6,33.024682540071);

\path [draw=color2, very thick]
(axis cs:0.7,32.3741425176901)
--(axis cs:0.7,33.024682540071);

\path [draw=color2, very thick]
(axis cs:0.8,32.3741425176901)
--(axis cs:0.8,33.024682540071);

\path [draw=color2, very thick]
(axis cs:0.9,32.3741425176901)
--(axis cs:0.9,33.024682540071);

\addplot [thick, color0, dashdotted, mark=square, mark size=2, mark options={solid,fill=none}]
table {%
0.1 31.175100696318
0.2 31.1032917283851
0.3 31.1888274176766
0.4 31.0772140643234
0.5 31.1548183324362
0.6 30.6584719623978
0.7 29.8970548957385
0.8 28.864479078957
0.9 27.8311129494567
};
\addlegendentry{DES}
\addplot [thick, color1, dotted, mark=*, mark size=2, mark options={solid,fill=none}]
table {%
0.1 90.571471188482
0.2 89.5150163164799
0.3 91.0235124271466
0.4 94.5992403029232
0.5 95.4257498481892
0.6 89.3662908393347
0.7 95.1513971312242
0.8 90.0082871869028
0.9 89.4487235853244
};
\addlegendentry{AoIS}
\addplot [thick, color2, dashed]
table {%
0.1 32.6994125288806
0.2 32.6994125288806
0.3 32.6994125288806
0.4 32.6994125288806
0.5 32.6994125288806
0.6 32.6994125288806
0.7 32.6994125288806
0.8 32.6994125288806
0.9 32.6994125288806
};
\addlegendentry{GES}
\end{axis}

\end{tikzpicture}
	\caption{Achieved average quadratic error vs. discount factor $\gamma$ for discounted error scheduler (DES), greedy error scheduler (GES) and age-of-information scheduler (AoIS). Out of five control loops, only one sub-system is allowed to transmit simultaneously due to single channel assumption, i.e., $N=5$ and $R=1$. Success probability of each transmission is $p_i = 0.9 ~ \forall i$. Vertical error bars represent 95\% confidence intervals for Monte Carlo simulations with 100 repetitions.}
	\label{fig:avgQE}
\end{figure}
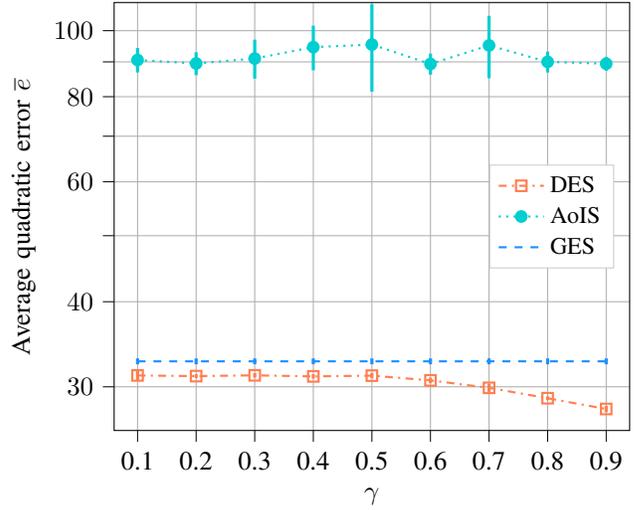

Next we plot our main performance metric $\overline{e}$ against increasing $\gamma$ in Fig.~\ref{fig:avgQE}. One can observe that the AoIS leads to much higher estimation error in average due to its unawareness of control system parameters. Therefore, one can say that AoI scheduling is not suitable for NCS scenarios. On the other hand, DES achieves steadily lower error than the GES and AoIS. In addition, the error is lower for increasing discount factor. Thus, we deduce that being far-sighted and valuing future costs, as done by the DES with higher $\gamma$, achieves better estimation accuracy than rushing immediate rewards as done by the GES or low $\gamma$ values.

\subsection{Discussion on selecting $M$}
One of the key design parameters for the proposed scheduling algorithm is the selection of $M$. For any given augmentation level $M$ and number of users $N$, the cardinality of the state space $\approxstatespace$ equals to $M^N$. Hence, the scalability of the approach suffers heavily from an increase in $M$. On the other hand, if a larger $M$ is selected, the approximating sequence approaches to the original MDP and may potentially lead to an improvement in performance. As a result, selection of $M$ is a trade-off between complexity and performance for the proposed approach.

\begin{figure}[t]
	\centering
\begin{tikzpicture}

\definecolor{color0}{rgb}{0.803921568627451,0.52156862745098,0.247058823529412}
\definecolor{color1}{rgb}{0.392156862745098,0.584313725490196,0.929411764705882}
\definecolor{color2}{rgb}{1,0.270588235294118,0}
\definecolor{color3}{rgb}{1,0.713725490196078,0.756862745098039}
\definecolor{color4}{rgb}{0.545098039215686,0,0.545098039215686}

\begin{axis}[
legend cell align={left},
legend style={at={(0.45,0.75)}, anchor=east, draw=white!80.0!black, font=\small},
tick align=outside,
tick pos=left,
x grid style={white!69.01960784313725!black},
xlabel={\(\displaystyle \gamma\)},
xmajorgrids,
xmin=0.06, xmax=0.94,
xtick style={color=black},
xtick={0,0.1,0.2,0.3,0.4,0.5,0.6,0.7,0.8,0.9,1},
xticklabels={0.0,0.1,0.2,0.3,0.4,0.5,0.6,0.7,0.8,0.9,1.0},
y grid style={white!69.01960784313725!black},
ylabel={Average quadratic error \(\displaystyle \overline{e}\)},
ymajorgrids,
yminorticks=true,
ymode=log,
ymin=24.6297566068359, ymax=87.893985439254,
ytick style={color=black},
ytick={30,40,50, 60, 70, 80},
yticklabels={\(\displaystyle {30}\),\(\displaystyle {40}\),\(\displaystyle {50}\), \(\displaystyle {60}\), \(\displaystyle {70}\), \(\displaystyle {80}\)}
]
\path [draw=color0, thick]
(axis cs:0.1,31.244578169236)
--(axis cs:0.1,31.7951896174141);

\path [draw=color0, thick]
(axis cs:0.2,32.1529247278822)
--(axis cs:0.2,33.3765013363425);

\path [draw=color0, thick]
(axis cs:0.3,32.3082646648544)
--(axis cs:0.3,33.6241934151854);

\path [draw=color0, thick]
(axis cs:0.4,37.7098745383975)
--(axis cs:0.4,41.254068837079);

\path [draw=color0, thick]
(axis cs:0.5,39.6104029763935)
--(axis cs:0.5,47.3598572307137);

\path [draw=color0, thick]
(axis cs:0.6,34.701859176352)
--(axis cs:0.6,85.018338674144);

\path [draw=color0, thick]
(axis cs:0.7,41.447769803227)
--(axis cs:0.7,47.4713671360369);

\path [draw=color0, thick]
(axis cs:0.8,35.5446853153133)
--(axis cs:0.8,41.4756314561126);

\path [draw=color0, thick]
(axis cs:0.9,31.839267148801)
--(axis cs:0.9,34.8510752568672);

\path [draw=color1, thick]
(axis cs:0.1,31.5982517310533)
--(axis cs:0.1,32.3001040728752);

\path [draw=color1, thick]
(axis cs:0.2,32.0114062295982)
--(axis cs:0.2,32.7512092000384);

\path [draw=color1, thick]
(axis cs:0.3,32.0479195350849)
--(axis cs:0.3,32.5981931404228);

\path [draw=color1, thick]
(axis cs:0.4,32.2441807725012)
--(axis cs:0.4,33.0151491893338);

\path [draw=color1, thick]
(axis cs:0.5,33.6310630069726)
--(axis cs:0.5,34.6872247586721);

\path [draw=color1, thick]
(axis cs:0.6,34.7060086697603)
--(axis cs:0.6,36.16962785165);

\path [draw=color1, thick]
(axis cs:0.7,33.4287600816247)
--(axis cs:0.7,34.5852015298971);

\path [draw=color1, thick]
(axis cs:0.8,31.4567728923953)
--(axis cs:0.8,33.2633552380687);

\path [draw=color1, thick]
(axis cs:0.9,29.3491840581262)
--(axis cs:0.9,30.6524744393718);

\path [draw=color2, thick]
(axis cs:0.1,30.740984548061)
--(axis cs:0.1,31.2272412085015);

\path [draw=color2, thick]
(axis cs:0.2,31.3111555643441)
--(axis cs:0.2,32.1091908664046);

\path [draw=color2, thick]
(axis cs:0.3,31.5135541915157)
--(axis cs:0.3,32.0296719919897);

\path [draw=color2, thick]
(axis cs:0.4,31.4013341121131)
--(axis cs:0.4,31.9694716786469);

\path [draw=color2, thick]
(axis cs:0.5,31.8892702335927)
--(axis cs:0.5,32.5042403617228);

\path [draw=color2, thick]
(axis cs:0.6,31.2639008265682)
--(axis cs:0.6,32.0224312938579);

\path [draw=color2, thick]
(axis cs:0.7,30.3059613009855)
--(axis cs:0.7,31.0301956931087);

\path [draw=color2, thick]
(axis cs:0.8,29.7279829929498)
--(axis cs:0.8,30.4299744315219);

\path [draw=color2, thick]
(axis cs:0.9,28.2550792478666)
--(axis cs:0.9,28.8350657429327);

\path [draw=color3, thick]
(axis cs:0.1,30.6969659622556)
--(axis cs:0.1,31.1778096996534);

\path [draw=color3, thick]
(axis cs:0.2,30.853500068648)
--(axis cs:0.2,31.2999593952654);

\path [draw=color3, thick]
(axis cs:0.3,31.0723682575964)
--(axis cs:0.3,31.6704246379489);

\path [draw=color3, thick]
(axis cs:0.4,31.0648790854533)
--(axis cs:0.4,31.6287200269302);

\path [draw=color3, thick]
(axis cs:0.5,31.0216843860737)
--(axis cs:0.5,31.8472652614912);

\path [draw=color3, thick]
(axis cs:0.6,30.2598371130303)
--(axis cs:0.6,30.7037556160463);

\path [draw=color3, thick]
(axis cs:0.7,29.7455014296433)
--(axis cs:0.7,30.4850317446494);

\path [draw=color3, thick]
(axis cs:0.8,28.6203866558041)
--(axis cs:0.8,29.1159865024869);

\path [draw=color3, thick]
(axis cs:0.9,27.5054033719458)
--(axis cs:0.9,27.9202024690427);

\path [draw=color4, thick]
(axis cs:0.1,30.684359974467)
--(axis cs:0.1,31.6658414181689);

\path [draw=color4, thick]
(axis cs:0.2,30.9019018311362)
--(axis cs:0.2,31.3046816256339);

\path [draw=color4, thick]
(axis cs:0.3,30.9478809231564)
--(axis cs:0.3,31.4297739121967);

\path [draw=color4, thick]
(axis cs:0.4,30.8430150555296)
--(axis cs:0.4,31.3114130731172);

\path [draw=color4, thick]
(axis cs:0.5,30.872807934486)
--(axis cs:0.5,31.4368287303864);

\path [draw=color4, thick]
(axis cs:0.6,30.3686363748133)
--(axis cs:0.6,30.9483075499822);

\path [draw=color4, thick]
(axis cs:0.7,29.4882093235807)
--(axis cs:0.7,30.3059004678963);

\path [draw=color4, thick]
(axis cs:0.8,28.6771790556013)
--(axis cs:0.8,29.0517791023127);

\path [draw=color4, thick]
(axis cs:0.9,27.5314772527112)
--(axis cs:0.9,28.1307486462022);

\addplot [thick, color0, solid, mark=*, mark size=2, mark options={solid,fill=none}]
table {%
0.1 31.5198838933251
0.2 32.7647130321124
0.3 32.9662290400199
0.4 39.4819716877382
0.5 43.4851301035536
0.6 59.860098925248
0.7 44.459568469632
0.8 38.5101583857129
0.9 33.3451712028341
};
\addlegendentry{DES M=15}
\addplot [thick, color1, dashed, mark=square, mark size=2, mark options={solid,fill=none}]
table {%
0.1 31.9491779019642
0.2 32.3813077148183
0.3 32.3230563377538
0.4 32.6296649809175
0.5 34.1591438828224
0.6 35.4378182607051
0.7 34.0069808057609
0.8 32.360064065232
0.9 30.000829248749
};
\addlegendentry{DES M=16}
\addplot [thick, color2, dashdotted, mark=triangle, mark size=2, mark options={solid,fill=none}]
table {%
0.1 30.9841128782812
0.2 31.7101732153743
0.3 31.7716130917527
0.4 31.68540289538
0.5 32.1967552976578
0.6 31.6431660602131
0.7 30.6680784970471
0.8 30.0789787122359
0.9 28.5450724953997
};
\addlegendentry{DES M=17}
\addplot [thick, color3, dotted, mark=diamond, mark size=2, mark options={solid,fill=none}]
table {%
0.1 30.9373878309545
0.2 31.0767297319567
0.3 31.3713964477726
0.4 31.3467995561918
0.5 31.4344748237824
0.6 30.4817963645383
0.7 30.1152665871464
0.8 28.8681865791455
0.9 27.7128029204942
};
\addlegendentry{DES M=20}
\addplot [thick, color4, loosely dotted, mark=triangle, mark size=2, mark options={solid,rotate=90,fill=none}]
table {%
0.1 31.175100696318
0.2 31.1032917283851
0.3 31.1888274176766
0.4 31.0772140643234
0.5 31.1548183324362
0.6 30.6584719623978
0.7 29.8970548957385
0.8 28.864479078957
0.9 27.8311129494567
};
\addlegendentry{DES M=25}
\end{axis}

\end{tikzpicture}
	\caption{Achieved average quadratic error vs. discount factor $\gamma$ for discounted error scheduler (DES). Each curve belongs to one of the five selected augmentation levels $M \in \{15, \, 16, \, 17, \, 20, \, 25\}$. Vertical error bars represent 95\% confidence intervals for Monte Carlo simulations with 100 repetitions.}
	\label{fig:Mevaluation}
\end{figure}
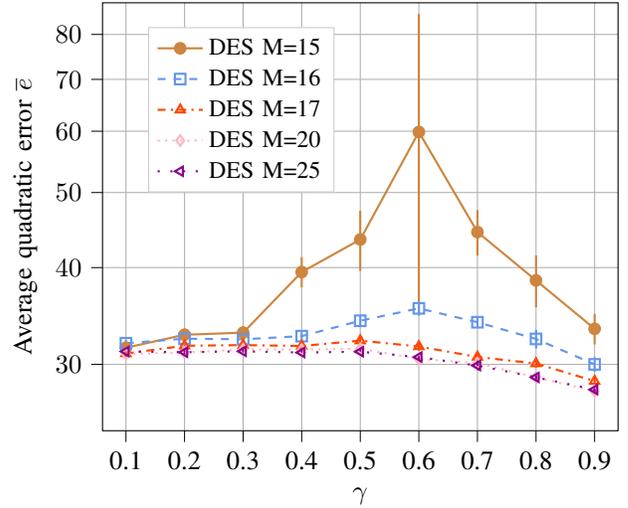

In order to investigate the impact of the augmentation level on the performance, we vary the design parameter $M$ and obtain the stationary optimal scheduling policy as described in Algorithm \ref{alg:valueiteration}. Fig. \ref{fig:Mevaluation} plots our key performance metric, the average quadratic error $\bar{e}$, over $\gamma$ for five different $M$ values, i.e, $M \in \{15, \, 16, \, 17, \, 20, \, 25\}$. If we look at the $M=15$ curve, we observe that it is outperformed by the other policies in terms of $\bar{e}$. This is a result of the relatively inaccurate cost functions due to the approximation. That is, since we limit the maximum cost to $g_i(\tilde{\Delta}^{(M)})$ for each class $i$, the approximating finite sequence does not represent the original MDP as accurate as when higher $M$ is selected. As a result, an increase in $M$ leads to a reduction in $\bar{e}$. However, the performance gain diminishes after a certain augmentation level which is evident especially from $M=20$ and $M=25$ curves. Their overlapping behavior justifies that there is almost no performance gain by introducing any additional complexity to the proposed scheduler design.

\begin{figure}[t]
	\centering
	\input{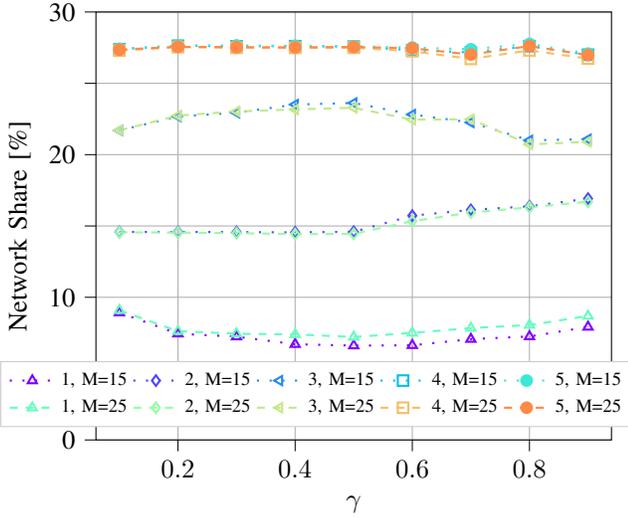}
	\caption{Illustration of network resources shares for different sub-system classes over 100 simulation runs. Numbering in the legend indicate the class $i$ with the respective system matrices $A_{1,2,3,4,5} = \{1.1, 1.3, 1.5, 1.7, 1.9\}$. Identical markers belong to the same class $i$. Dotted and dashed lines illustrate the average network resource share granted to $i$ by the discounted error scheduler (DES) with augmentation levels $M=15$ and $M=25$, respectively. 95\% confidence intervals are too small to be displayed.}
	\label{fig:networkSharesM}
\end{figure}

Note that $\bar{e}$ shows a concave behavior for $M=15$. This can be explained with the help of Fig. \ref{fig:networkSharesM} which plots average network shares of individual classes $i$ over $\gamma$ for $M=15$ and $M=25$. Both policies follow very similar trajectories for the same class $i$ except for the most stable one with the system matrix $A_1 = 1.1$. This follows from overlooking potential high future costs of class $i = 1$ around $\gamma = 0.6$ due to low bounds we enforce on $g_i(\tilde{\Delta}^{(M)})$ and prioritizing $i=3$ instead. This effect is reversed by increasing $\gamma$ further. For $M=25$ this occurs less dramatically than for $M=15$.

\section{CONCLUSIONS AND FUTURE WORKS}
\label{sec:conclusions}
Age-of-Information has been recently used in both control and communication research for data scheduling in multi-user scenarios. In the context of NCS, it serves as an interface between information update frequency and control performance. In this work, we studied the centralized scheduling problem for multiple networked control loops sharing a wireless channel with random packet loss. We were able to formulate the scheduling problem as a cost minimization problem over an infinite horizon. We have used AoI as the states of the MDP and obtained a stationary scheduling policy as a function of states. We have shown that by employing control dependent age penalty functions and solving the infinite horizon cost minimization problem, we can reduce error further than the schedulers existing in the literature. 
%

\bibliographystyle{IEEEtran}
\bibliography{references}

\end{document}